\documentclass[cits]{PoS}
\usepackage{epsfig}
\usepackage{graphicx}
\usepackage{amsmath}
\usepackage{amsfonts}
\usepackage{amssymb}
\usepackage{bm}
\usepackage{times}
\usepackage{mathptmx}
\usepackage{fontenc}
\newcommand{\gammab}{\mbox{\boldmath$\gamma$}}
\newcommand{\nablab}{\mbox{\boldmath$\nabla$}}

\title{Quintessence, Neutrino Masses and Unification of the Dark Sector}

\ShortTitle{Quintessence, Neutrinos and Dark Sector}

\author{\speaker{Gennady Y. Chitov} \\
        Department of Physics, Laurentian University, Sudbury, ON, P3E 2C6 Canada\\
        E-mail: \email{gchitov@laurentian.ca}}

\abstract{The origin of the neutrino mass, dark matter (DM), and dark energy
(DE) are among the most challenging problems of fundamental physics. We address
these questions from analyses of the models where the neutrino mass is
generated via Yukawa coupling to the quintessence field which represents the
DE. It has been shown in a recent work on the toy model with a single Dirac
fermion coupled to the quintessence that by choosing parameters of the DE
potential to match the present DE density, the model allows to lock the
neutrino mass at $m \sim 10^{-2}~eV$ and yields consistent estimates for other
parameters of the Universe. To include the DM component into this framework we
propose here to add the right-handed Majorana term(s) into Lagrangian with the
quintessence-generated mass. As a results, the DE field is responsible for
generation of the masses of the light active Majorana neutrinos as well as of
the heavy sterile neutrinos. The latter are natural DM candidates. }

\FullConference{The XXth International Workshop  High Energy Physics and Quantum Field Theory     \\
          September 24 - October 1,  2011 \\
         Sochi, Russia}

\begin{document}

\section{Introduction}

It appears now that consensus has being reached on our understanding of the
composition of the Universe and fundamental issues demanding extensions of the
standard model of particle physics. Roughly 75\% of the Universe is comprised
of dark energy (DE), about 22\% of it is comprised of dark matter (DM), and
only approximately 3\% corresponds to the contribution of the conventional
(baryonic) matter. The present stage evolution of the Universe is governed by
the dominant DE contribution, and the Universe experiences an accelerating
expansion. The nature of the both DE and DM is still unknown, and this is
probably one of the greatest mysteries of modern physics \cite{DE}. The origin
of neutrino mass is also a fundamental problem for particle physics as well as for
cosmology \cite{Dolgov08,Les06}. Current upper
limits on the sum of neutrino masses from cosmological observations are of the
order of 1eV, while neutrino oscillations give a lower bound of roughly 0.01 eV
\cite{Les06}. Even if  at least \textit{ad hoc}, minimalistically and
intellectually unsatisfactory, the DE can be accounted for by adding a
cosmological term to fit current observations, the existence of neutrino masses
and DM particles definitely requires revisions of the standard model.
There is a vast literature and a great interest in these issues. Due to space
constraints we address the reader to ``a short list''
\cite{DE,Dolgov08,Les06,ShaposhnikovNuMSM,Shaposhnikov,Kayser03} where more references,
especially on earlier work could be found.

In this paper we summarize and advance further our recent work \cite{Us11}.
It stems from the earlier proposal by Fardon \textit{et al} \cite{Fardon04}.
In the approach \cite{Us11} the origin of the neutrino mass is due to the dark energy
in the form of quintessence. Both the DE density and the neutrino mass are
determined by a single mass scale of the DE potential, and consistent
estimates for other parameters of the Universe are obtained.

Here we propose how to incorporate  the ``third element of the puzzle'', i.e, the
DM component of the Universe within a common framework. For this end we analyze a quite
straightforward extension of the ``toy-model'' of a single Dirac fermion considered
in \cite{Us11} by adding a right-handed Majorana fermion whose mass is due to the
Yukawa interaction with the quintessence field. The Dirac-Majorana action for a single
generation of neutrinos is known to be equivalent to two massive Majorana fermions.
One of these, the (light) neutrino with mass $m_a$ decreasing in time can be
identified with the conventional (active) neutrino, while the (heavy) sterile
neutrino with growing mass $m_s$ is a natural candidate for the DM particle.
The results and estimates we report below demonstrate that the theory is viable and
consistent in unifying neutrino masses and the dark sector of the Universe.

\section{Toy Model: Dirac Fermion Coupled to the Ratra-Peebles Quintessence \cite{Us11}}

The standard methods of general relativity and finite-temperature quantum field
theory \cite{Kapusta} were applied for fields in the flat
Friedmann-Lema\^itre-Robertson--Walker (FLRW) Universe. The scale factor $a(t)$
is governed by  the Friedmann equations. The DE is modeled by the bosonic
scalar field (quintessence) in the FLRW metric with the Euclidian action:
\begin{equation}
\label{SEBa}
    S_B= \int_0^\beta d \tau \int a(t)^3 d^3 x ~\Big[
    \frac12 (\partial_\tau \varphi)^2 +\frac{1}{2 a^2}
    (\nabla \varphi)^2 + U(\varphi) \Big]~,
\end{equation}
where $\varphi= \varphi(\mathbf{x}, \tau)$ and $\hbar=c=k_B=1$. The analysis is
restricted to the Ratra-Peebles quintessence potential:
\begin{equation}
\label{RP}
    U(\varphi)=\frac{M^{\alpha+4}}{\varphi^{\alpha}}~,
\end{equation}
with $ \alpha>0$. In this toy model a single massless neutrino is mimicked by the the Dirac Euclidian action
\begin{equation}
\label{SEDa}
     S_F=\int_0^\beta d \tau \int a(t)^3 d^3 x ~\bar{\psi}(\mathbf{x}, \tau)
     \Big(\gamma^o \frac{\partial}{\partial \tau}-\frac{\imath}{a}
    \gammab \cdot \nablab \Big) \psi(\mathbf{x}, \tau).
\end{equation}
The fermion and boson sectors of the Euclidian Lagrangian are interacting via the Yukawa coupling
term:
\begin{equation}
\label{LYuk}
    \mathcal{L}_{YD}= g \varphi \bar{\psi} \psi~.
\end{equation}
Note that the Ratra-Peebles potential $U(\varphi)$ does not have a nontrivial minimum,
so the generation of the fermion mass is due to the coupling between the quintessence and
fermions resulting in a nonzero
average of the bosonic field: $m=g \langle \varphi \rangle$. (In the following the dimensionless Yukawa
coupling is set $g=1$.) Evaluation of the grand partition function in the saddle-point
approximation \cite{Us11} yields the density of the thermodynamic potential
\begin{equation}
\label{OmTree}
    \Omega( \varphi)= U( \varphi  )+
    \Omega_F( \varphi)~,
\end{equation}
calculated at $\varphi=\langle \varphi \rangle$, where $\Omega_F$ is the
standard density of the thermodynamic potential of the free Dirac field
\cite{Kapusta}. The average of the scalar field  is determined by minimization:
\begin{equation}
\label{Min}
    \frac{\partial \Omega(\varphi)}{\partial \varphi}
    \Big\vert_{\varphi=\langle \varphi \rangle}=U^\prime (\varphi)+ g \rho_{ch}= 0~,
\end{equation}
where we introduce the  chiral fermionic condensate density:
\begin{equation}
\label{Trhos}
    \rho_{ch} =\frac{2 m}{ \pi^2 \beta^2} \int_{\beta m}^\infty
    \frac{(z^2-(\beta m)^2)^{\frac12} }{e^z+1}dz~.
\end{equation}
Analysis of the extrema of the thermodynamic potential (\ref{Min}) reveals three phases of the
coupled model.

\begin{figure}
\includegraphics[width=0.35\textwidth]{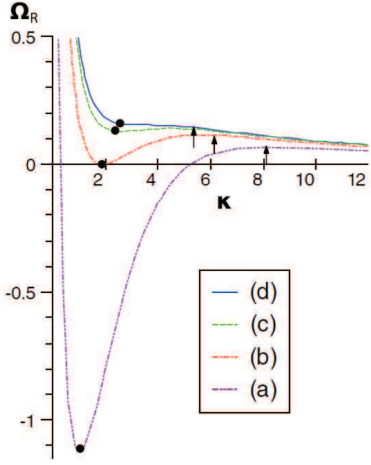}
\caption{ The dimensionless density of the thermodynamic potential $\Omega_R
\equiv \Omega / M^4$ as a function of the dimensionless parameter $\kappa
\equiv g \varphi /T$. The thermodynamically stable solutions of
Eq.~({\protect\ref{Min}}) indicated by the large dots correspond to the minima
of the potential (curve a). The arrows indicate unstable solutions,
corresponding to the maxima of the potential. The curve b is calculated at the
temperature of metastability $T_m$. At lower temperature $T_c$ the local
minimum becomes an inflection point of the potential (curve d), and the system
gets unstable. This is a point of spinodal decomposition. Adapted from
\cite{Us11}. } \label{IaOm}
\end{figure}

\textsl{1). Stable phase, $T>T_m$:}  At high temperatures the equation
(\ref{Min}) has two nontrivial solutions. The root indicated with a large dot
in Fig.~\ref{IaOm} (case a) corresponds to a global minimum of the potential.
It is a thermodynamically \textit{stable state}. In this phase $\Omega <0$, so
the pressure $P=- \Omega$ is positive. In the high-temperature phase the
fermionic mass is small:
\begin{equation}
\label{msmall}
    m \approx M \Big( \sqrt{6 \alpha} \frac{M}{T}  \Big)^{\frac{2}{\alpha+2}}
    \propto T^{-\frac{2}{\alpha+2}}~,
\end{equation}
The fermionic contribution is dominant and the thermodynamic potential behaves
to leading order as that of the ultra-relativistic fermion gas. Two masses in
this model demonstrate opposite temperature dependencies. The quintessence
field with the conventionally defined mass
\begin{equation}
\label{mphi}
    m_\phi^2=  \frac{\partial^2 U(\varphi)}{\partial \varphi^2}
    \Big\vert_{\varphi=\langle \varphi \rangle}~,
\end{equation}
is ``heavy'' at high temperatures:
\begin{equation}
\label{mphiUP}
    m_\phi \approx \sqrt{\frac{\alpha+1}{6}} T~,~~ M/T \ll 1~,
\end{equation}
however its mass decreases together with the temperature. In contrary, the fermionic
mass $m$ monotonously increases with decreasing temperature.

Another non-trivial root of (\ref{Min}) corresponds to a thermodynamically \textit{unstable state}
(maximum of $\Omega$ indicated with an arrow in Fig.~\ref{IaOm}. There is a trivial third root
$\varphi = \infty$ corresponding to the ``doomsday''
vacuum with a vanishing dark energy. In the high-temperature phase the doomsday vacuum $\Omega=0$ is
a \textit{metastable state} of the Universe.

%
%

\textsl{2). Metastable phase, $T_m < T<T_c$:} At the temperature of
metastability $T_m$  the thermodynamic potential has two degenerate minima
$\Omega(\langle \varphi \rangle )=P(\langle \varphi \rangle)=\Omega(\infty)=0$.
This is shown in Fig.~\ref{IaOm} (case b). Below this temperature the two
minima of the potential exchange their roles. The root indicated ny a large dot
becomes a metastable state with $\Omega(\langle \varphi \rangle)>0$, i.e., with
the negative pressure $P(\langle \varphi \rangle)<0$, while the stable state of
the system corresponds to the true stable vacuum of the Universe
$\Omega(\infty)=P(\infty)=0$. See Fig.~\ref{IaOm} (case c). The state in the
local minimum $\Omega(\langle \varphi \rangle)$ is analogous to a metastable
supercooled liquid. We disregard here the exponentially small sphaleron
contributions of the fermions hopping from the metastable state $\Omega(\langle
\varphi \rangle)$ into the vacuum state $\Omega(\infty)=0$.

\textsl{3). Phase transition at $T=T_c$ (spinodal decomposition):}
At the critical temperature
\begin{equation}
\label{Tc}
   T_c \approx  \frac{M}{\Delta_c} ~
\end{equation}
two extrema of the potential merge into an inflection point. This is shown in
Fig.~\ref{IaOm} (case d). Here we define
\begin{equation}
\label{Dcr}
    \Delta_c =
    \Big( \frac{\sqrt{2}}{\alpha \pi^{3/2}} \nu^\nu e^{-\nu} \Big)^{\frac{1}{\alpha+4}}~,
    ~~ \nu = \alpha + \frac52 ~.
\end{equation}
From the viewpoint of equilibrium thermodynamics  the model must undergo a
first-order (discontinuous) phase transition and reach its thermodynamically
stable (at $T<T_c$) phase corresponding to the doomsday vacuum $\Omega(\varphi=
\infty)=P(\varphi= \infty)=0$. At the critical point the sound velocity and the
compressibility vanish, thus the phase transition occurs through
\textit{spinodal decomposition}. During this transition the fermionic mass
given at the critical point by
\begin{equation}
\label{mc}
    m_c  \approx \frac{\nu}{\Delta_c} M~
\end{equation}
and the quintessence mass
\begin{equation}
\label{mphic}
    m^c_\phi \approx
    \sqrt{\alpha(\alpha+1)} \Big(\frac{\Delta_c}{\nu} \Big)^{\frac{\alpha+2}{2}}M
\end{equation}
both jump from the above values to their values in the vacuum state
$m=\infty$ and $m_\phi=0$.

\begin{figure}
\includegraphics[width=0.8\textwidth]{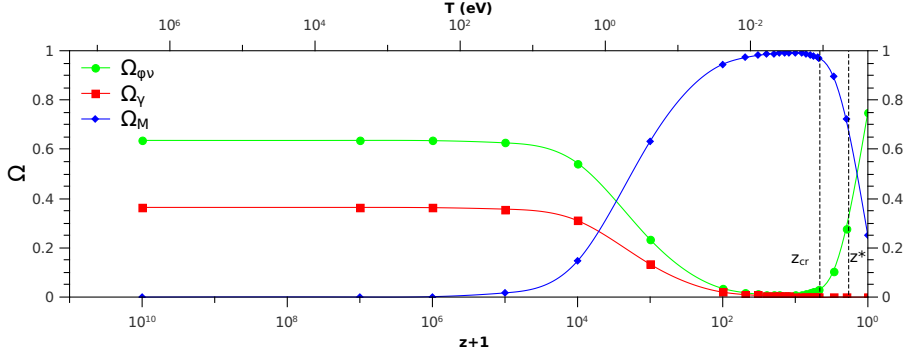}
\caption{ Relative energy densities plotted up to the current redshift (temperature, upper axis):
$\Omega_{\varphi\nu}$ -- coupled DE and neutrino contribution; $\Omega_{\gamma}$ --
radiation; $\Omega_{M}$ -- combined baryonic and dark matters. Parameter
$M=2.39 \cdot 10^{-3}$~eV ($\alpha=0.01$), chosen to fit the current densities,
determines the critical point of the phase transition $z_c \approx 3.67$.
The crossover redshift $z^* \approx 0.83$ corresponds to the  point where the
Universe starts its accelerating expansion. Adapted from \cite{Us11}.}
\label{Omega}
\end{figure}

It is quite a nontrivial problem to analyze the time evolution of the Universe
towards the new equilibrium vacuum state. In our work \cite{Us11} we abandoned
completely the thermodynamic or kinetic considerations below $T_c$ and resort
to the solution of the equation of motion
\begin{equation}
\label{EqMotion}
    \ddot{\varphi}+ 3 H \dot{\varphi}+\frac{\partial \Omega}{\partial
    \varphi}=0~.
\end{equation}
together with the Friedmann equations. The parameters found at the critical
point were used as the initial conditions for the dynamics at $T<T_c$. The
numerical solution for the quintessence field $\varphi(t)$ from the critical
point to the present time oscillates quickly (with the period $\tau \sim
10^{-27}$ Gyr) around the smooth (``mean value'') solution $\bar \varphi (t)$.
The latter is found analytically as
\begin{equation}
\label{fMD}
  \bar \varphi = \varphi_c \cdot
     \Big( \frac{1+ z_c}{1+z}\Big)^{\frac{3}{\alpha+1}} ~.
\end{equation}
Relating the mean values with the physically relevant
observable quantities, the key results are obtained analytically. To connect the only
physical scale of the model $M$ with the observables, we set the current density of
the quintessence field coupled to the fermions to be equal to
the observable value of the dark energy, i.e., roughly $3/4$ of the critical density:
\begin{equation}
\label{DEnow}
    \rho_{\varphi\nu, \mathrm{now}}=\rho_{DE} \approx
    \frac34 \cdot \frac{3 H_0^2}{8 \pi G}
    \approx 31 \cdot 10^{-12} ~\mathrm{eV}^4~,
\end{equation}
and the combined baryonic and dark matter $ \rho_{M, \mathrm{now}} \approx
\frac14 \cdot \frac{3 H_0^2}{8 \pi G}~$. The exponent of the quintessence
potential $\alpha $ is then the only parameter which can be varied. It is found
that to get a realistic estimate for the present-time neutrino mass $m(t)=\bar
\varphi(t)$, one needs a small $\alpha$. For instance, at $\alpha =0.01$
$m_{\mathrm{now}} \approx 0.27$ eV. The relative energy densities $\Omega_\#
\equiv \rho_\#  /\rho_{\mathrm{tot}}$ as functions of redshift (or temperature)
are plotted in  Fig.~\ref{Omega}. There we also indicate the critical point
parameterized by the redshift $z_{\mathrm{c}}$ and the crossover point $z^*$.
The latter is defined as the redshift at which the Universe starts its
late-time acceleration, i.e., where $w_{\mathrm{tot}} \equiv P_{\mathrm{tot}} /
\rho_{\mathrm{tot}}=-\frac13$. For the present time we find $ w_{\mathrm{tot}}
\approx -\frac34$.

%
%

Being successful in relating the neutrino mass to the DE, this toy model have
two major flaws. First, it does not give a clue about (heavy) DM particles.
Secondly, a more serious problem is that in order to get a small neutrino mass
we need a small $\langle \varphi \rangle$. This makes the mass too sensitive to
the local temperature and thus the existence of the fundamentally identical
particles in the Universe very problematic. As we will see in the next section
the seesaw mechanism in the extended model allows to resolve these two problems
on the same footing.

\section{Extension of the toy model: Seesaw, Active and Sterile Majorana Neutrinos, Dark Energy}

To elucidate key points a theory needs to unify all three elements -- neutrino,
DM and DE within the same framework, we extend minimally the above toy model.
For simplicity we take only one flavor of neutrinos. In addition to the Dirac
mass term in the Lagrangian (\ref{LYuk}) written in terms of the left- and
right-handed components of the Dirac field $\psi$ as $\mathcal{L}_{YD}= m_D
\bar{\psi}_R \psi_L~+~h.c.$, let us assume the existence of the right-handed
Majorana neutrinos with the mass term $m_R$:
\begin{equation}
\label{LMaj}
    \mathcal{L}_{YM}= \frac12 m_R \bar{\psi}_R^c \psi_R~+ h.c.~,
\end{equation}
where the charge conjugation is defined as $\psi_R^c =- \imath \gamma^2
\psi_R^*$. The extensions of the standard model to include the Majorana mass
term have been known for quite a while (for reviews and original references,
see, e.g., \cite{ChengLi84,Bilenky87}), and more recent work pertinent to the
present context in \cite{ShaposhnikovNuMSM,Shaposhnikov,Kayser03}). We take the
standard assumption from the earlier work
(\cite{ChengLi84,Bilenky87,ShaposhnikovNuMSM,Shaposhnikov,Kayser03} and more
references there in), that the Dirac mass is due to the Yukawa coupling to the
neutral Higgs scalar, i.e., $m_D =g_D \langle H \rangle$. We propose here that
the origin of the Majorana neutrino mass is due to the coupling to the dark
energy (quintessence), i.e., $m_R =g \langle \varphi \rangle$. To the best of
our knowledge this was not studied before. The proposal closest to the present
one is put forward by Shaposhnikov and co-workers
\cite{ShaposhnikovNuMSM,Shaposhnikov} in the framework of the so-called
$\nu$MSM extension of the standard model, where the right-handed neutrinos get
their masses through the coupling to the inflaton field.\footnote{It is quite
plausible that the inflaton and quintessence represent the same physical field
 analyzed at the different regimes of the Universe evolution. In this sense the current proposal might
 turn out to be equivalent to the one of \cite{Shaposhnikov}.}

Treatment of the fermionic action with the Dirac and Majorana terms is a
well-known (seesaw) problem \cite{ChengLi84,Bilenky87}. The action can be
brought to the diagonal form of the two Majorana fermions with positive masses:
\begin{equation}
\label{Msa}
    m_{s,a}= \frac12 \Big( \sqrt{m_R^2+ 4m_D^2} \pm m_R  \Big)~.
\end{equation}
Taking the Dirac mass as a fixed parameter, we treat the model of two Majorana fermions
coupled to the quintessence along the lines of the problem with a single Dirac fermion \cite{Us11}.
In the saddle-point approximation we get the thermodynamic potential (\ref{OmTree}) with the
fermionic contribution from two Majorana fields and minimization equation
\begin{equation}
\label{MinMaj}
    U^\prime (\varphi)+ \frac12 \frac{g}{\sqrt{m_R^2+ 4m_D^2}}
    \Big( m_s \rho_{ch}(m_s)- m_a  \rho_{ch}(m_a)\Big)= 0~,
\end{equation}
Let us now set the physically acceptable range of the model's parameters. We
set the quintessence Yukawa coupling $g=1$ (this can always be done by
rescaling of $M$ \cite{Us11}) and take the mass of the active light Majorana
neutrino at present time $m_a = 10^{-2}$ eV. Since we don't know the Yukawa
coupling $g_D$ between the Dirac and Higgs fields ($\langle H \rangle =174$
GeV), we put $m_D$ in the range from the temperature of electroweak scale where
the Higgs bosons acquire masses, i.e., $m_D \sim T_{EW} \sim 10^2$ GeV to the
scale of light leptons $m_D \sim 1$ MeV. Then the current mass of the heavy
(sterile) Majorana neutrino is actually given by the quintessence $m_s \approx
m_R \approx m_D^2/m_a$ and lies within $10^{15}- 10^{5}$ GeV. Matching then the
current Ratra-Peebles potential to the DE density (cf. Eq.~(\ref{DEnow})) we
get the range $M \sim 5 \cdot 10^{2}-5$ eV. For reader's convenience we put the
model's parameters together with some other quantities in
Table~\ref{Parameters}. Out of curiosity we also added there an entry with an
ultralight Dirac mass $m_D=1$ KeV.

At high temperatures $T> T_{EW}$ when $m_D=0$ the model is in a stable phase
whose parameters (up to a factor 2) coincide with that of the above case of a
single Dirac fermion \cite{Us11}. The mass of active neutrino $m_a=0$, while
the sterile neutrino is light and relativistic $m_s=m_R  \approx \mathcal{O}(1)
\cdot M \Big(M/T  \Big)^{2/(\alpha+2)}$.

At $T< T_{EW}$ behavior of the model with Majorana neutrinos demonstrates some
new qualitative features comparatively to the case \cite{Us11} due to:
\textit{(i)} the new mass scale $m_D$ which sets in below $T_{EW}~$;
\textit{(ii)} decreasing $m_a$ with growing quintessence. The Dirac mass brings
in a new critical temperature
\begin{equation}
\label{TcD}
    T_{cD}= \frac25 m_D ~,
\end{equation}
below which the sign of the second term in the minimum equation (\ref{MinMaj})
is always negative and no nontrivial solution for decreasing quintessence
potentials exists.\footnote{For large $m_D \sim T_{EW}$ this opens an
interesting opportunity for the future work to explore trial DE potentials
other than the Ratra-Peebles one.}

\begin{table}
\begin{tabular}{|l|l|l|l|}
  \hline
   $~~~~~~~~~~~~~~T$  & $T \geq T_{EW} \sim 10^2$ GeV  & $T=T_c \approx 0.4 m_D$ & $T=T_{now}$ \\
   $M$ & $m_R=m_s  \sim M \big(\frac{M}{T} \big)^{2/(\alpha+2)}$ & $m_R^c \approx 6.42M$ & $m_R \approx m_s \approx  m_D^2/m_a$ \\
  $m_D$ & $m_a=m_D=0$  & $m_{s,a}^c \approx m_D \pm \frac12 m_R^c$ & $m_a$ ~-~fixed\\
  \hline
  $5 \cdot 10^2$ eV &$m_R ~~~~~\leq 10^{-3}$ eV & $3$ KeV & $10^{15}$ GeV \\
  $10^2$ GeV      & $m_s~~~~~ \leq 10^{-3}$ eV & $10^2$ GeV &  $10^{15}$ GeV\\
  ~ & $m_a~~~~~ ~~~0$   & $10^2$ GeV &  $10^{-2}$ eV\\
  \hline
  $5 $ eV &$m_R ~~~~~\leq 10^{-6}$ eV  & $30$ eV & $10^{5}$ GeV \\
  $1$ MeV      & $m_s~~~~~ \leq 10^{-6}$ eV  & $1$ MeV &  $10^{5}$ GeV\\
  ~ & $m_a~~~~~ ~~~0$   & $1$ MeV &  $10^{-2}$ eV\\
  \hline
  $0.3 $ eV &$m_R ~~~~~ \leq 10^{-9}$ eV  & $2$ eV & $10^{8}$ eV \\
  $1$ KeV      & $m_s~~~~~ \leq 10^{-9}$ eV & $1$ KeV &  $10^{8}$ eV\\
  ~ & $m_a~~~~~ ~~~0$   & $1$ KeV &  $10^{-2}$ eV\\
  \hline
\end{tabular}
\caption{Neutrino masses at different temperatures ($\alpha=1$) for three choices of
the Dirac mass $m_D$. The current active neutrino mass is set $m_a=10^{-2}$ eV and the
quintessence potential scale $M$ is chosen to match the current DE density.
All the parameters in the table are defined in the text.}
\label{Parameters}
\end{table}

The evolution of the model below $T_{EW}$ towards the critical point of the
first order phase transition is qualitatively similar to that shown in
Fig.~\ref{IaOm}. The critical temperature where the maximum and minimum of the
thermodynamic potential merge into an inflection point is found along with the
critical value of the mass $m_R$:
\begin{equation}
\label{TcMaj}
  T_c \approx T_{cD}
     \Big( 1+ \zeta_\alpha \Big(  \frac{M}{m_D}\Big)^2  \Big)~,~~~m_R^c \approx \vartheta_\alpha M~,
\end{equation}
where the numbers $\zeta_\alpha \sim \vartheta_\alpha \sim \mathcal{O}(1)$. For
$\alpha =1$ we find $\zeta_1 \approx 2.86$ and  $\vartheta_1 \approx 6.42$. As
one can see from Table~\ref{Parameters} in the physically interesting range of
parameters $T_c \approx T_{cD}$, so the critical temperature is determined in
fact solely by the Dirac mass $m_D$ (compare to Eq.~(\ref{Tc})), while the
quintessence at $T_c$ is controlled by the Ratra-Peebles scale $M$, similar to
Eq.~(\ref{mc}). The quintessence field being heavy at high temperatures $m_\phi
\sim T$, becomes lighter with cooling of the Universe, reaching $m_\phi^c \sim
M$ at $T_c$, and vanishes below $T_c$. We have checked that the sound velocity
and the compressibility vanish at $T_c$, thus this the spinodal decomposition
point and the transition occurrs via rapidly growing density fluctuations below
$T_c$.

\section{Discussion: Realistic Extension of the Standard Model, Dark Matter and Dark Energy}

It appears that the above model of Majorana neutrinos interacting with the
quintessence and getting their masses from DE is consistent, and it encourages
for further work on it. But we have to stress that below the phase transition
the model is probed so far only by consistency checks between various
parameters ($\rho_{DE},~m_a,~m_s$) and observables. Analysis of the equation of
motion (\ref{EqMotion}) for the quintessence field and/or by using more general
methods of kinetics of the first-order phase transition is warranted.

However, the model itself needs to be more realistic. It seems to us that the
so-called $\nu$MSM  \cite{ShaposhnikovNuMSM,Shaposhnikov} is the right model
which has already properties needed. This is an extension of the standard model
where three additional right-handed Majorana gauge singlets for each of the
neutrino flavors are added. With the model's available fitting parameters it is
possible \cite{ShaposhnikovNuMSM} to get along with the three light active
neutrinos also the three sterile neutrinos with consistent abundances, masses,
and life times.\footnote{In addition this model also provides a mechanism for
baryon asymmetry of the Universe.} One of the sterile neutrinos with the life
time comparable to the age of the Universe has a mass in the KeV range and is
an excellent ``warm'' DM candidate, while the other short-lived heavy neutrinos
have their masses in the GeV range and are decaying DM particles. The results
presented here allow us to anticipate that using the quintessence field for
generating the right-handed Majorana massive terms (\ref{LMaj}) in this
extended standard model would also allow to obtain the observable DE density
and the slow-rolling quintessence dynamics with $w_\varphi \equiv P_\varphi  /
\rho_\varphi \approx -1$. This study is currently work in progress.

\acknowledgments
I am thankful to my co-authors T. August, A. Natarajan and especially to T. Kahniashvili
for collaboration on our paper \cite{Us11}. I also thank A. Dolgov and M. Shaposhnikov for
correspondence. I am very grateful to O. Chkvorets for many helpful
discussions and bringing important references to my attention.
The author acknowledges financial support from the Natural
Science and Engineering Research Council of Canada (NSERC) and the Laurentian
University Research Fund (LURF).


\end{document}